\documentclass[preprint,showpacs,preprintnumbers,amsmath,amssymb,
nofootinbib]{revtex4}
\usepackage{graphicx}
\usepackage{dcolumn}
\usepackage{bm}

\newcommand{\nn}{\nonumber}

\newcommand{\be}{\begin{eqnarray}}
\newcommand{\ee}{\end{eqnarray}}
\catcode`\@=11
\def\lsim{\mathrel{\mathpalette\@versim<}}
\def\gsim{\mathrel{\mathpalette\@versim>}}
\def\@versim#1#2{\vcenter{\offinterlineskip
\ialign{$\m@th#1\hfil##\hfil$\crcr#2\crcr\sim\crcr } }}
\catcode`\@=12

\begin{document}
\vspace{2cm}
\preprint{KANAZAWA-09-06}

\title{
Anomaly Induced Dark Matter Decay \\and\\
PAMELA/ATIC Experiments}

\author{Hiroki Fukuoka}
\author{Jisuke Kubo}
\author{Daijiro Suematsu}

\affiliation{
 Institute for Theoretical Physics, Kanazawa
University, Kanazawa 920-1192, Japan
\vspace{3cm}
}

\begin{abstract}

\vspace{1cm} 
The cosmic ray data of PAMELA/ATIC
may be explained  by dark matter decay
with a decay rate
$\tau_{DM}^{-1}\sim 10^{-26}~\mbox{sec}^{-1}
~\sim 10^{-45}~\mbox{eV}$,
an energy scale which could not be understood within the framework
of the standard model or its simple supersymmetric extension.
We  propose anomaly induced dark matter decay to
exponentially suppress the decay rate, and apply
to a supersymmetric extension  of the Ma's inert Higgs model
of the radiative seesaw mechanism for neutrino masses.
In this model  the lightest right-handed neutrino $\psi_N$
and the lightest neutralino $\chi$  can
 fill the observed necessary dark matter relic,
 and we find that $\psi_N$ can decay into $\chi$
 through anomaly with a right order of decay rate, 
 emitting only leptons.
 All the emitted poistrons are right-handed.
 
 \end{abstract}

\maketitle

\section{Introduction}
Recent astrophysical observations \cite{wmap} and 
neutrino oscillation experiments \cite{oscil}
require an extension of the standard model (SM) so as to include
 dark matter   as well as to incorporate a generation mechanism for small neutrino 
masses.
At the moment, however, we know about dark matter only a little;
the constraint on its mass and abundance, but nothing about 
its detailed feature is known. Consequently, there are many 
consistent models for 
dark matter. Representative possibilities may be summarized 
as follows: \\
(i) Dark matter is a stable thermal relic, so that its relic abundance and
annihilation cross section are strongly related to each other.
The lightest neutralino in supersymmetric models 
with the conserved $R$-parity 
is a well studied example of this category \cite{susydm}. 
Another well motivated example may be a stable neutral 
particle  in the 
radiative seesaw scenario \cite{Ma:2006km}, which is an alternative model of 
the seesaw mechanism to generate  small neutrino masses \cite{seesaw}.
In fact, there exits similar models and the nature of
the DM candidates in these models has been studied \cite{scdm,cdmmeg,
fcdm,ncdm,ext}.\\ 
(ii) Dark matter consists of  multiple components \cite{multidm}. If some of them are 
unstable, dark matter can contain thermal components as well as 
non-thermal ones which can be produced by the decay of unstable components.
In this case the dark matter relic abundance at present and the annihilation
cross section of the dominant component need not to be related.\\
(iii) Dark matter is not stable and is decaying with a very long
lifetime  \cite{Takayama:2000uz}-\cite{Shirai:2009fq}. 

In any case it will be crucial for the study for going beyond the SM to know 
which class the true dark matter model belongs to.
Since the above mentioned dark matter models 
predict different signals for 
the annihilation and/or decay of dark matter in the Galaxy,
it may be possible to use 
the data from these observations to distinguish the dark matter 
models \cite{mindep}.
The positron excess in the recent PAMELA observation \cite{pamela} and 
ATIC data of $e^++e^-$ flux \cite{atic} are such examples.
PAMELA data show a hard positron excess compared with the background but 
no antiproton excess, while 
ATIC data show the excess of $e^++e^-$ flux at regions of 300-800 GeV. 
A lot of works have been done to explain these data within the framework
of  dark
matter models (see, for instance,  \cite{positron} and references therein).
However, the observed positron flux requires much larger
annihilation cross section or much larger dark matter density than 
the ones needed for the explanation of WMAP data.
The required 
large factor in the latter feature is parametrized as a boost factor 
in the references\footnote{The necessary enhancement for the s-wave
annihilation can be partly covered by
the Sommerfeld effects \cite{Hisano:2003ec} or others \cite{enhance}.}.
So, it seems very difficult to 
give a natural explanation for the boost factor for the type (i) dark matter. 

In this paper, following
 \cite{Ma:2006uv}, 
 we consider a supersymmetric extension of the radiative 
seesaw model for the neutrino mass 
to understand  the data obtained by the
PAMELA and ATIC experiments.
The radiative seesaw model is  attractive in two respects:
(a) The non-vanishing
small neutrino mass
and  the presence of a dark matter candidate are closely related through
a discrete symmetry $Z_2$.
(b) The dark matter candidate in this model couples only with 
leptons but not quarks. This feature is favorable for
the above mentioned PAMELA and ATIC data. 
However, a large boost
factor still has to be introduced to explain the observed positron flux  in this
model \cite{Bi:2009md,Cao:2009yy,Chen:2009mf}. 
In our  supersymmetric 
extension of the model this problem is overcome
as follows.
There are two kinds of  stable neutral 
particles corresponding to  two discrete symmetries,
$R$ and $Z_2$, where $R$ is the $R$ parity
in supersymmetric theories, and $Z_2$ is mentioned above.
If one of these
discrete symmetries is broken,
the heavier one can decay to the lighter one.
We propose that this breaking can be induced by anomaly 
 \cite{Banks:1991xj,Banks:1995ii,ArkaniHamed:1998nu} to realize
an exponentially suppressed decay rate of the heavier dark matter.
It should be noted that the smallness of this decay rate is a crucial 
ingredient for the explanation of the observed $e^++e^-$ flux. 
Moreover, due to the very nature of the model, 
only lepton pairs can be produced through the dark matter decay.
We show that both data of PAMELA and ATIC can be described 
well simultaneously in this scenario. 
The model for dark matter proposed in this paper  gives a concrete 
realistic example of type (ii). 

\section{Anomaly induced dark matter decay}

The stability of  the dark matter  is usually ensured by
an unbroken discrete symmetry $Z$.
If the discrete symmetry is broken, the dark matter can decay.
The   preferable decay modes depend
on how $Z$ is broken.
However, its life time will be too short $\tau_{DM} \sim (8/\pi) m_{NDM}
\simeq 10^{-24} $ sec for $m_{DM} \simeq 1$ TeV, unless the $Z$ breaking
is extremely weak 
\cite{Takayama:2000uz}-\cite{Shirai:2009fq}. 
Such suppression may occur if $Z$ is
broken  by  GUT or Planck scale 
physics 
\cite{Hamaguchi:2008ta}-\cite{Shirai:2009fq}.

Here we would like to suggest an alternative 
suppression mechanism which 
is based on the observation that
if a symmetry, continuous or discrete,  is anomalous,
 then  non-perturbative effects can generally  
 induce non-invariant terms, like
quark masses in QCD.
Although the discrete symmetry
$Z$ in question  can be anomaly free with respect to the SM gauge group,
it can be anomalous at high energy when imbedded into a larger
discrete group, because heavy particles
can contribute to discrete anomalies \cite{Ibanez:1991hv,Banks:1991xj}. 
If the discrete symmetry is anomalous at high energy,
 non-perturbative effects can produce 
 $e^{-b S} \Phi^n$ in the superpotential 
 \cite{Banks:1991xj,Banks:1995ii,ArkaniHamed:1998nu}, 
 which is $Z$ invariant. This is  because  the dilaton superfield 
$S$ transforms inhomogeneously
under the anomalous $Z$, where  $\Phi$ is  a generic chiral super field, and
$b$ is a certain real number
(see also \cite{Araki:2007ss,Araki:2008ek}).
Below the Planck scale, where the dilaton is assumed 
to be stabilized at a vacuum
expectation value of $O(1)$, the factor  $e^{-b <S>}$ can work
as a suppression factor for the noninvariant  product $\Phi^n$.

Let us estimate the size of the suppression. 
To this end, consider a (chiral) $Z_N$ symmetry in a gauge theory
based on the gauge group $G$ and assume $Z_N$  is anomalous. Then 
the Jacobian $J$ of the path integral measure 
corresponding to the $Z_N [G]^2$ anomaly can be written as 
\cite{Araki:2006mw,Araki:2007ss,Araki:2008ek}
\be
 J &=&
  \exp\left(-\frac{2\pi i}{N}\Delta Q \int d^{4}x~2{\cal A}(x)
        \right),  {\cal A}(x) =\frac{1}{64\pi^{2}}
        \epsilon^{\mu\nu\rho\sigma}
       \ F_{\mu\nu}^a F_{\rho\sigma}^a,
       \label{jac}
\ee
where $F_{\mu\nu}^a$ is the field strength for $G$. Since 
the Pontryagin index $\int d^{4}x~{\cal A}(x)$ is 
an integer, $\Delta Q =0$ mod $ N/2$  means anomaly freedom
of $Z_N$. In the anomalous case, we have $\Delta Q =k/2$ with
an integer $k < N$. (So, $\Delta Q/N$=1/4 for an 
anomalous $Z_2$, for instance.)
This anomaly can be cancelled by the Green-Schwarz
mechanism \cite{Green:1984sg}, 
which defines the transformation property of the dilaton
supermultiplet $S=(\varphi+i a, \psi_S,F_S)$,
where $\varphi$ ($a$) is  the dilaton (axion) field, and they couple
to the gauge field as
\be
{\cal L}_F &=&
      -\frac{\varphi}{4} \ F_{\mu\nu}^a F^{a\mu\nu}
       - \frac{a}{8}\epsilon^{\mu\nu\rho\sigma}
       \ F_{\mu\nu}^a F_{\rho\sigma}^a.
\ee
To cancel the anomaly (\ref{jac}), the axion $a$ has to transform 
according to $a \to a -(1/2\pi) (\Delta Q/N)$.
Therefore, the $Z_N$ charge of $\exp(-bS)$ becomes $C$ if
$b = 4 \pi^2 C/\Delta Q$. Since $<\varphi>=1/g^2\simeq O(1)$
at the Planck scale, 
the expression $\exp(-bS)$ would then yield a suppression factor $SF$
such as
\be
SF &\simeq& \exp(-4 \pi^2 C/\Delta Q),\nn
\\
(SF)^2  &\simeq & 10^{-55}, 10^{-69}, 
  10^{-86}~~~\mbox{for}~
  C/\Delta Q =8/5, 2, 10/4,
        \label{sf}
        \ee
where $C$ and $2\Delta Q$ are defined modulo  $N$
\footnote{Since $b>0$, $C+N$ should appear instead of $C$ for a negative $C$.
}.

Do we need such a big suppression? According to
\cite{Ibarra:2008jk,Ishiwata:2008cu}, to explain
the PAMELA/ATIC data, 
the decaying dark matter
 should decay with a life time of  $\sim 10^{26}$ sec, which corresponds
to  a decay width $\Gamma_{NDM} \sim  10^{-54}\times(1 \mbox{TeV})
\sim   10^{-70}\times(1 \mbox{TeV}) (M_{\rm PL}/1 \mbox{TeV})
\sim  10^{-86}\times(1 \mbox{TeV}) (M_{\rm PL}/1 \mbox{TeV})^2$,
where we have assumed that the decay is induced by dimension four (three)
operators for the first (third) expression. (The second one 
could appear accidentally.) The precise suppression needed depends, of course,
on the details of the model. But it is clear that one needs 
a huge suppression factor
for the decaying dark matter,
 if one would like to explain
the PAMELA/ATIC data within the framework of particle physics
 \cite{Hamaguchi:2008ta,Arvanitaki:2008hq,Shirai:2009fq}. 
 It is also clear that the existence of  such a small number can not 
be explained 
in a low energy theory.

\section{The Model: Radiative see-saw and  dark matter candidates}
Here we would like to supersymmetrize the model of \cite{Ma:2006km}.
(An first attempt has been made in \cite{Ma:2006uv}.)
We assume the $R$ parity invariance as usual.
So, we have $R\times Z_2$ discrete symmetry at low energy.
The matter content of the model with their quantum number
is given in Table I.
$L$, $ H^u, H^d$ and $\eta^u, \eta^d$
stand for $SU(2)_L$ doublets
supermultiplets of the leptons, the MSSM  Higgses and the inert 
Higgses, respectively.
Similarly, $SU(2)_L$ singlet
supermultiplets of the charged leptons and right-handed neutrinos are denoted by
$E^c$ and $N^c$. $\phi$ is an additional neutral Higgs supermultiplet 
which is needed
to generate neutrino masses radiatively. 
$\Sigma$ and $\sigma$ are also  additional neutral Higgs 
supermultiplets which are
 needed to derive the superpotential (\ref{superP1}) from
 a $Z_4$ invariant one.

\begin{table}[t]
\begin{center}
\begin{tabular}{|c|cccccccc|cc|} \hline
 & $L$ & $E^c$ & $N^c $&$H^u$&$H^d$& $\eta^u $ 
 & $\eta^d$ & $\phi $ &  $\Sigma$ & $\sigma$
  \\ \hline
 $R\times Z_2 $ 
 & $(-,+)$  &  $(-,+)$  &  $(+,-)$
 & $(+,+)$&  $(+,+)$
 &  $(-,-)$ & $(-,-)$& $(-,-)$ &  $(+,+)$ &  $(+,+)$ 
   \\ \hline
    $Z_4$ 
 & $0$  &  $0$  &  $-1$
 & $0$&  $0$
 &  $1$ & $1$& $-1$ & $2$ & $0$ 
   \\ \hline
$Z_2^L$ 
 & $-$  &  $-$  &  $-$
 & $+$&  $+$
 &  $+$ & $+$& $+$& $+$& $+$
   \\ \hline
\end{tabular}
\caption{The matter content and
the quantum number. $Z_{2L}$ is a discrete lepton number.
$Z_2$ is a subgroup of $Z_4$, which is assumed to be anomalous
and spontaneously broken by VEV
of $\Sigma$ and $\sigma$ down to $Z_2$.
}
\end{center}
\end{table}

We first consider 
a  $R\times Z_2$ invariant
superpotential below. Later on, using $\Sigma$ and $\sigma$, we will describe a 
possibility  to 
obtain it from a $R\times Z_4$ invariant one:
\be
W &=& W_4+W_2,
\label{superP1}
\ee
where
\be
W_4
&=&
Y_{i}^{e} L_{i} E_{i}^c  H^d
+Y_{ij}^{\nu } L_{i} N_{j}^c \eta^u
+\lambda_u \eta^u H^d \phi+
\lambda_d \eta^d H^u \phi
+\mu_H H^u H^d,
\label{w4}\\
W_2 &=& \frac{(M_N)_{ij}}{2}N_i^c N_j^c,+\mu_\eta \eta^u \eta^d+
\frac{1}{2}\mu_\phi\phi^2.
\label{w2}
\ee
The Yukawa couplings of the charged leptons $Y_{i}^{e}$
can be assumed to be diagonal without loss of generality.

Soft-supersymmetry breaking terms are necessary to generate neutrino masses
radiatively. For the relevant Higgs sector they are given by
\be
{\cal L}_{SB}
&=&
-m_{\eta^u}^2\hat{\eta}^{u\dag} \hat{\eta}^u-
m_{\eta^d}^2\hat{\eta}^{d\dag} \hat{\eta}^d
-m_{\phi}^2 \hat{\phi}^\dag \hat{\phi}
-(B_\eta \hat{\eta}^u \hat{\eta}^d+h.c.)\nn\\ 
& &-
(\frac{1}{2}B_\phi\hat{\phi}^2+h.c.)+
(A_u \lambda_u \hat{\eta}^u \hat{H}^d \hat{\phi}+
A_d \lambda_d \hat{\eta}^d \hat{H}^u \hat{\phi} +h.c.),
       \label{LSB}
\ee
where the hatted field is
the scalar component of the corresponding
superfield. The $B$ and $A$ soft terms are responsible 
for the radiative generation of the neutrino masses.
We assume that $\hat{\eta}^u, \hat{\eta}^d$ and  $\hat{\phi}$
do not acquire vacuum expectation values. 

To calculate
the one-loop neutrino mass matrix, we treat the $B$ terms 
as insertions. Then we find
a one-loop diagram with one insertion of
 $B_\eta \hat{\eta}^u \hat{\eta}^d$,
which mixes $\hat{\eta}^u$ and  $\hat{\eta}^d$.
Correspondingly,  we define the approximate
mass eigenstates $\eta_0^\pm$ (the neutral
component of $\eta$) as
\be
\left(\begin{array}{c}
\eta^u_0 \\ \eta^{d}_0 \end{array}\right) &=&
\left(
\begin{array}{cc}
\cos\theta & \sin\theta \\
-\sin\theta &\cos\theta \end{array}\right)~
\left(\begin{array}{c}
\eta^+_0 \\
\eta^-_0 \end{array}\right),
\ee
where
\be
\tan2\theta &=&-\frac{2 m_{ud}^2}{m_{uu}^2-m_{dd}^2},~
 m_{\pm}^2 =
\frac{1}{2}\left\{m_{uu}^2+m_{dd}^2\pm
[(m_{uu}^2-m_{dd}^2)^2+4 m_{ud}^4]^{1/2}\right\},
       \label{t2t}
\ee
with
\be
m_{uu}^2 &=&
\mu_\eta^2+m_{\eta^u}^2
-\frac{1}{2}M_z^2\cos 2\beta 
+\frac{1}{2}\lambda_u^2v^2\cos^2\beta,\\
m_{dd}^2 &=&
\mu_\eta^2+m_{\eta^d}^2
+\frac{1}{2}M_z^2\cos 2\beta
+\frac{1}{2}\lambda_d^2v^2\sin^2\beta,\\
m_{ud}^2 &=&\frac{1}{2}\lambda_u\lambda_d v^2 \cos\beta\sin\beta,
       \label{mud}
\ee
and  $\tan\beta=v_u/v_d~,~v^2=v_u^2+v_d^2~,~
M_z =(g_2^2+g^{\prime 2})v^2/4$.
Neglecting
higher order insertions we  obtain the neutrino mass matrix at  one loop:
\be
({\bf M}_\nu)_{ij} &=&
\frac{1}{16\pi^2} Y^\nu_{il} U_{l k} M_k
U^T_{km} Y^\nu_{jm}B_\eta \sin 2\theta
\left[-\cos^2 \theta I(m_+,m_+,M_k)\right.\nn\\
& &\left.+\sin^2\theta I(m_-,m_-,M_k)
+\cos 2\theta I(m_+,m_-,M_k)\right],
       \label{mnu}
\ee
where $U$ is a unitary matrix defined by
$(U^T M_N U)_{ik}=M_k\delta_{ik}$
\footnote{$M_1 < M_{2,3}$ is assumed, and we denote
$M_1$ by $m_{NDM}$ later on.}, and
\be
& &I(m_a,m_b,m_c) \nn\\
& &=\int_0^1 dx \int_0^{1-x} dy
[m^2_a x+m^2_b y+m^2_c (1-x-y)]^{-1}\nn\\
& &=
\frac{m_a^2 m_c^2 \ln(m_a^2/m_c^2)+
m_b^2 m_c^2 \ln(m_b^2/m_c^2)+
m_b^2 m_a^2 \ln(m_b^2/m_a^2)
}{(m_a^2-m_b^2)
(m_b^2-m_c^2)(m_c^2-m_a^2)}.
\ee
As we can see from (\ref{t2t}), (\ref{mud}) and (\ref{mnu}), 
the neutrino masses are proportional to
$B_\eta$ and $\lambda_u \lambda_d$ at the lowest order, because $\sin 2\theta
\propto \lambda_u \lambda_d$. So, the neutrino masses can be controlled by these
parameters along with the Yukawa couplings $Y^\nu_{il}$, the
masses of the inert Higgses and right-handed neutrinos.

There are many candidates for the dark matter in this model \cite{Ma:2006uv}.
The lightest combination
of each row in Table II could be a dark matter.
\begin{table}[t]
\begin{center}
\begin{tabular}{|c|c|c|} \hline
 $R\times Z_2\times Z_2^L $ & Bosons &Fermions \\ \hline
  $(-,+,+)$  &  
  & $ \psi_{h^u},\psi_{h^d}, {\tilde Z}, {\tilde \gamma}$\\ \hline
  $(+,-,+)$ &     &  $\psi_{\eta^u}, \psi_{\eta^d}, \psi_\phi$ 
           \\ \hline
  $(-,-,+)$ &  $\hat{\eta}_0^u,\hat{\eta}_0^d,
  \hat{\phi}$ &   \\ \hline
  $(+,-,-)$ &  $\hat{N}$'s &  \\ \hline
  $(-,-,-)$ &  & $\psi_N $'s 
     \\ \hline
    $(-,+,-)$  & ${\hat \nu}_L$'s &
       \\ \hline
\end{tabular}
\caption{The dark matter candidates. The $(+,+,-)$ candidates 
are dropped, because
they are the left-handed neutrinos.}
\end{center}
\end{table}
But there can exist only three types of dark matter including the
left-handed neutrinos depending on which discrete symmetry guarantees
their stability.
We assume
that the first right-handed neutrino $\psi_{N_1}$ is the lightest  one
among $\psi_{N}$'s and denote it by  $\psi_{N}$ (its mass is denoted by 
$m_{NDM}$). So, $\psi_{N}$
and  the lightest neutralino $\chi$
( its mass is denoted by $m_{\chi DM}$)
are the dark matter candidates. Both have an odd $R$ parity, so that 
one of them can be the stable dark matter, while the other one
is the decaying dark matter, if $Z_2$ is broken.

Following \cite{Griest:1988ma,Griest:1989zh},  we have computed the
thermally averaged cross section for the annihilation of two $\psi_N$'s
and that of two $\chi$'s by expanding the corresponding relativistic
cross section $\sigma$ in powers of their relative velocity,
and we  have then computed the relic densities
$\Omega_{NDM}$ and $\Omega_{\chi DM}$.
 We have assumed that the SM particles are the only ones
which are lighter than $\psi_N$ and  $\chi$,
so that we have used the SM degrees of freedom  at the decoupling, i.e.
 $g_*=106.75$.  We have found that,
 for the given interval of the dark matter masses, i.e.,
 $1~\mbox{TeV}~\lsim m_{NDM}\lsim 3~\mbox{TeV}$
  and  $0.2~\mbox{TeV}~\lsim m_{\chi DM}\lsim 0.5~\mbox{TeV}$,
  there is an enough parameter space
  in which $(\Omega_{NDM}+\Omega_{\chi DM}) h^2\simeq 0.11$ is satisfied.
  In the next section we let $\psi_N$ decay into $\chi$, while emitting
high energy positrons. As it is clear from the superpotential (\ref{w4}),
$\psi_N$ can not decay into the quarks,  because $\eta$'s do not
couple to the quarks.

\section{Decaying right-handed neutrino dark matter and 
PAMELA/ATIC Data}

As long as the discrete symmetry $R\times Z_2$ is unbroken, there are two CDM
particles in the present model. One finds that the $R\times Z_2[SU(3)_C]^2$ and 
$R\times Z_2[SU(2)_L]^2 $ anomalies are canceled with the matter content
given Table 1 \footnote{We do not consider the $R\times Z_2[U(1)_Y]^2$
and mixed gravitational  anomalies, because they do not give us
useful informations.}.
Our  assumption is that $Z_4$
is anomalous and spontaneously broken to its subgroup $Z_2$.
Note that $Z_4$ forbids $W_2$ in (\ref{w2}) while $W_4$ is allowed.
Therefore, we have to produce it from an additional sector.
This situation can be realized as follows. Consider the $Z_4$ 
invariant superpotential
including the SM singlet $\Sigma$ and $\sigma$ given in Table 1:
\be
W_\sigma &=&
\xi \sigma  +m_\sigma\sigma^2+\lambda_\sigma\sigma^3
+ \lambda_\Sigma \sigma \Sigma^2+\lambda_\mu\sigma H^u H^d\nn\\
& &+m_\Sigma \Sigma^2
+\left(~ \frac{(\lambda_N)_{ij}}{2}N_i^c N_j^c+\lambda_\eta \eta^u \eta^d+
\frac{1}{2}\lambda_\phi\phi^2\right)\Sigma .
\label{Ws}
\ee
The superpotential (\ref{Ws}) serves for $\Sigma$ and $\sigma$ to develop VEVs,
and consequently, $Z_4$ is spontaneously broken to $Z_2$, 
producing effectively the superpotential (\ref{w2}).
The true stable dark matter is the lightest one
which has an odd parity of $R$. 
In the following discussion we assume that $\psi_N$ is heavier than $\chi$.
Since the ATIC data are indicating that the mass of the decaying 
dark matter particle is preferably
heavier than $O(1)$ TeV, all the
superpartners should be  heavier than   $O(1)$ TeV
if $m_{\chi DM} > m_{NDM}$.
It is, therefore,  more welcome for $\psi_N$ to be  the 
decaying dark matter, because
a heavy $\psi_N$ means a heavy $\eta$ Higgs, which is desirable to suppress
FCNC processes such as $\mu \to e + \gamma$.

As one can find, $Z_4$ is anomalous: $\Delta Q=1~\mbox{mod}~N/2(=2)$
\footnote{For the Green-Schwarz cancellation to work, the $Z_4[SU(3)_C]^2$
anomaly has to be matched to  $Z_4[SU(2)_L]^2$ anomaly.
To realize this we introduce, for instance, a pair
of ${\bf 3}$ and $\overline{{\bf 3}}$ of $SU(3)_C$ with the $Z_4$
charge  one. Their mass can be obtained from 
$<\Sigma> {\bf 3} \times \overline{{\bf 3}}$.}.
Consequently, 
the suppression coefficient
$b$ of (\ref{sf}) can take values
\be
b &=& \frac{4\pi^2 C}{\Delta Q}  =4\pi^2\times \frac{C~~(\mbox{mod}~4)}
{1~~(\mbox{mod}~2)}, 
\label{b}
\ee
where $C$ is the charge of
$\exp(-bS)$.
We assume that the non-perturbative effect
can generate $R$ invariant, but $Z_4$ violating  operators.
At $d=3$ there is only one operator
$\eta^u L$ which is even under $R$, and has the $Z_4$ charge one.
So we focus on $\eta^u L$:
\be
W_b &=&  \mu_{bi}\eta^u L_i \quad \mbox{with}~
\mu_{bi}=\rho_i M_{\rm PL}e^{-b \langle S\rangle},
\label{Wb}
\ee
where $\rho_i$ are dimensionless couplings.
Since   $\langle F_S/\varphi\rangle \sim m_{3/2}$  and 
$\langle\varphi\rangle\sim O(1)$, 
the superpotential $W_b$ induces a soft-supersymmetry breaking term
\be
{\cal L}_b &=& B_{bi}   \hat{\eta}^u \hat{L}_i \quad \mbox{with}~
B_{bi}= w  \rho_i M_{\rm PL} m_{3/2}e^{-b}
\label{Lb}
\ee
at the Planck scale, where is $w$ a dimensionless constant.
Since the $Z_4$ charge of $\eta$ is $1$,
the charge of $\exp(-bS)$ has to be $-1~\mbox{mod}~4$, and then
$$ b =4\pi^2\times (\cdots 7/3,~11/5,~11/7,~ 7/5,~ 1\cdots),$$ 
which could give a huge suppression factor.

\begin{figure}[t]
\begin{center}
\includegraphics[width=10cm]{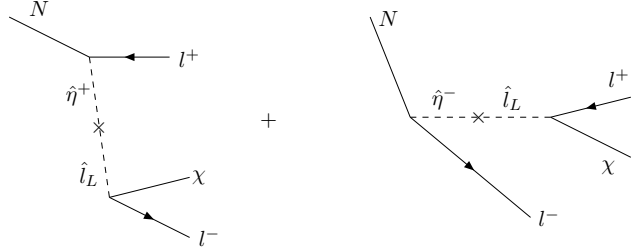}
\caption{\label{graph1}\footnotesize The diagrams contributing
to the $\psi_N$ decay. 
We assume that $\chi$ is the pure bino.
The emitted positrons are all right-handed.
Similar diagrams exist
because of the mixing of $\eta^u$ with $E^c$.
The amplitude is suppressed by $e^{-b} m_f/m_\eta$,
so that we do not consider them.
}
\end{center}
\end{figure}

With this observation we proceed with our discussion.
The tree diagrams contributing to the $\psi_N$ decay are shown 
in Fig.~\ref{graph1},
where we have assumed that $\chi$ is the pure bino.
We do not take into account the tree diagrams which
exist due to the mixing
of $\psi_{\eta^u}$ and $\psi_{e_f^c}$, because these diagrams 
are suppressed by $m_f/m_\eta$,
where $m_f$ is the lepton mass.
So, in the lowest order approximation only dimension two
operators in the $B$ soft-breaking sector exist.
At the lowest order, $\psi_N$ can decay only into 
the leptons along with a $\chi$. ($R$ parity violating
operator $L H^u$ allows the decay into the quarks, too.)
The differential decay width is given by
\be
\frac{d \Gamma_{e^+}}{d E} &=&
\frac{m_{NDM}^4}{768 \pi^3}x^2 \left(1-\frac{z^2}{1-2x}\right)^2
\Big[~A_1(1-2x -z^2)+2 A_1 (1-x) (1+2 \frac{z^2}{1-2x})\nn\\
& &+6 A_2 z+6 A_3 (1-2x)~\Big],
\label{Dg}
 \ee
 and the total decay width is
 \be
\Gamma_{e^+ T} &=&\tau_{NDM}^{-1}=\frac{m_{NDM}^5}{12288\pi^3}\Big\{~(1-z^2)
[  (A_1+A_3)(1-7 z^2-7 z^4+z^6)\nn\\
& &+4 A_2 z (1+10 z^2+z^4) ]+
24z^2[-(A_1+A_3)z+2 A_2 (1+z^2) \ln z]~\Big\},
\label{Tg}
 \ee
where
\be
z &=& \frac{m_{\chi DM}}{m_{N DM}} <1,\qquad x =\frac{E}{m_{N DM}} < (1-z^2)/2,
\label{z}
\ee
\be
A_1 &=&2g^{\prime 2}\sum_{i,j}  \left | Y_{j 1}^* B_i 
\right|^2\frac{1}{\tilde{m}_L^4 m_\eta^4} , \qquad
A_3 =2g^{\prime 2}\sum_{i,j} \left |
Y_{i 1}^* B_j \right|^2\frac{1}{\tilde{m}_L^4 m_\eta^4},
\label{a1} \\
A_2 &=& -g^{\prime 2}\sum_{i,j}
\left [( Y_{j 1} B_i ^*)(Y_{i 1}^* B_{j})+h.c. \right]
\frac{1}{\tilde{m}_L^4\tilde{m}_{\eta}^4},
\label{a2}
\ee
where $g' ~(\simeq 0.345)$ is the $U(1)_Y$ gauge coupling constant
(the  bino is assumed to be $\chi$).
$j$ runs over the negatively charged leptons, and
$i$ stands for a positively charged lepton in 
(\ref{a1}) and (\ref{a2}).  We have assumed that
all the scalar partners of the left-handed superpartners $\hat{l}_L$
have the same mass $\tilde{m}_L$.
The positron can come
from the decay of the anti-muons and anti-taus. 
In the following calculations, however,
we assume that the energy spectrum of the  positron coming from the anti-muon
and tau does not differ very much from that of the direct production
of the positron.
So, we also sum over $i=e^+, \mu^+,\tau^+$ 
in (\ref{a1}) and (\ref{a2}) to obtain 
$d \Gamma_{e^+}/d E$. At this order, all the emitted positrons are right-handed,
as one can see from Fig.~\ref{graph1}.

Before we calculate the positron spectrum, we briefly consider
the suppression factor we need for our case.
Assuming that $Y_{ij}\sim 1$ and
 that all the $\rho_i$ in $B$'s in (\ref{Lb}) are of the same size, we obtain
 \begin{equation}
 \tau_{NDM} \sim \left(\frac{ \mbox{TeV}}{m_{NDM}}\right)
 \left( \frac{m_\eta^2 \tilde{m}_L^2}{m_{NDM}^3 m_{3/2}} \right)^2
 \left(  \frac{m_{NDM}}
 {M_{\rm PL}/10^{16}}
 \right)^2 (\rho_{\tau} \omega)^{-2} \left(  10^{-79} e^{2b} \right) 
  \times 10^{26}  ~~\mbox{sec}.
 \end{equation}
 So,  we have a right order of $\tau_{NDM}$
with $b=4\pi^2(7/3)$ which gives a suppression factor of
$10^{-80}$ (see (\ref{b})).

\begin{figure}[t]
\begin{center}
\includegraphics[width=10cm]{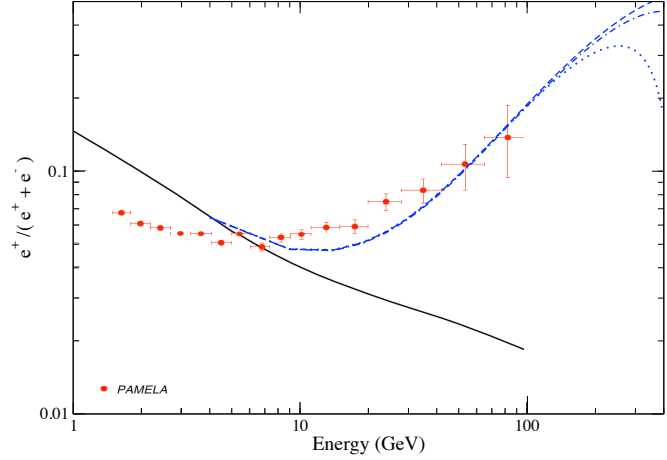}
\caption{\label{pamela}\footnotesize
$(e^+/(e^++e^-)$ versus the positron energy $E$.
The blue lines are the predictions of the model,
where we have used: $z=1/5\mbox{(dashed)}$, $1/6\mbox{(dot-dashed)}$, 
$1/5\mbox{(dotted)}$, $\tau_{NDM}(0.11/\omega_{NDM}
 h^2)=1.4\mbox{(dashed)}$, 
$2.0\mbox{(dot-dashed)}$, $3.0\mbox{(dotted)}~\times 10^{-26}~\mbox{sec}$,
$m_{NDM}= 2.0\mbox{(dashed)}$, $1.5\mbox{(dot-dashed)}$, 
$1.0\mbox{(dotted)}~\mbox{TeV}$.
The red points are the PAMELA data \cite{pamela}, 
where the  predictions  are  written over the
figure 4 of the PAMELA paper \cite{pamela}.
The solid line is the background published in  \cite{pamela}, and it agrees 
with the one calculated from
(\ref{p1})-(\ref{p3}) without the primary source of the positron.}
\end{center}
\end{figure}

\begin{figure}[t]
\begin{center}
\includegraphics[width=12cm]{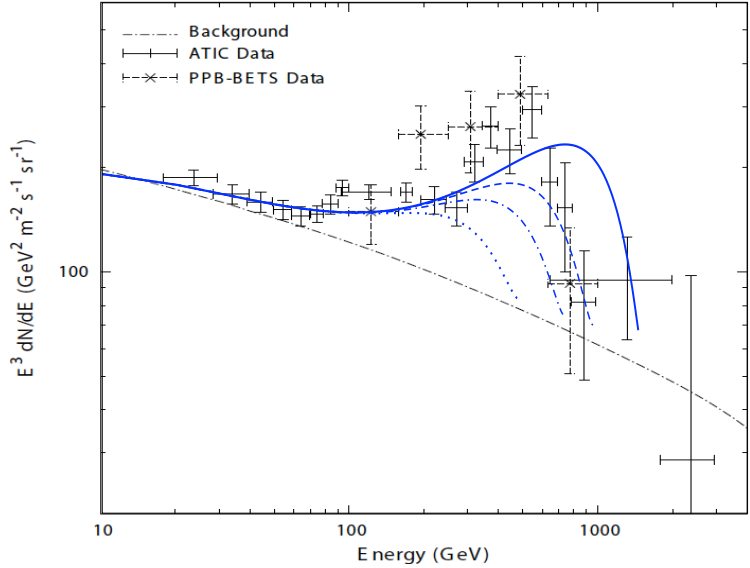}
\caption{\label{atic2}\footnotesize
The differential energy  spectrum scaled by $E^3$.
The solid, dashed, dot-dashed and dotted blue lines are the predictions
of the model. 
The  parameter values used here are the same as for Fig.~\ref{pamela}.
The solid blue line is calculated with  $z=1/10$,
$\tau_{NDM}(0.11/\omega_{NDM} h^2)=0.94
\times 10^{26}~\mbox{sec}$ and $m_{NDM}=3~\mbox{TeV}$.
The predictions  are  written over the
figure 3 of the ATIC paper \cite{atic}, where the PPB-BETS data
\cite{Torii:2008xu} are also plotted.
The black dashed line is the background
presented in  \cite{atic}.
The normalization factor $N_\phi = 0.76$ in (\ref{p1})-(\ref{p3}) is so chosen
that the background computed from (\ref{p1})-(\ref{p3})
agrees with the black dashed line at low energy.
}
\end{center}
\end{figure}

 Now we come to compute the positron spectrum:
\be
f_{e^+}(E) &=&\int_E^{E_{\rm max}} d E'G_{e^+}(E,E')~
 \frac{d \Gamma_{e^+}(E')}{d E'},
 \label{fe}
\ee
where $E_{\rm max} = (m_{NDM}^2-m_{\chi DM}^2)/2 m_{NDM}$, 
$d \Gamma_{e^+}(E)/d E=(\tau_{NDM})^{-1}
d n_{e^+}(E)/d E$, and we vary $\tau_{NDM}$ freely to fit
the data.
The positron Green's function $G_{e^+}$ 
of \cite{Hisano:2005ec} can be approximately written 
as \cite{Ibarra:2008jk}
\be
 G_{e^+}(E,E') &\simeq &\left(\frac{\Omega_{NDM}h^2}{0.11}\right)
 \frac{10^{16}}{E^2} 
 \exp [ a+b(E^{\delta-1}-E^{\prime \delta -1})   ]~~\mbox{cm}^{-3}~\mbox{s},
 \ee
 where  $a,b,\delta$ depend on the diffusion model
 \cite{Moskalenko:1997gh,Delahaye:2007fr,Ibarra:2008jk}. 
 Here we use those of the MED
 model \cite{Delahaye:2007fr}: $a=-1.0203, b=-1.4493, \delta=0.70$, 
and we have assumed that
 except for the normalization factor $\Omega_{NDM}h^2/0.11$
 the decaying dark matter $\psi_N$ has the same density profile in our galaxy
 as the NFW profile \cite{Navarro:1995iw}.
 The background differential  flux for each species are \cite{Baltz:1998xv}
\be
\Phi^{\rm prim. bkg}_{e^-}(E) &=&N_{\phi}0.16 E^{-1.1}\left[
1+11 E^{0.9}+3.2 E^{2.15}\right]^{-1},
\label{p1}\\
\Phi^{\rm sec. bkg}_{e^-}(E) &=&N_{\phi}0.7 E^{0.7}\left[
1+11 0E^{1.5}+600 E^{2.9}+580 E^{4.2}\right]^{-1},
\label{p2}\\
\Phi^{\rm sec. bkg}_{e^+}(E) &=&N_{\phi}4.5 E^{0.7}\left[
1+650 E^{2.3}+1500 E^{4.2}\right]^{-1}
\label{p3}
\ee
in the units in $[\mbox{GeV}~\mbox{cm}^2~\mbox{s}~\mbox{sr}]^{-1}$,
where  the energy $E$ is in the units in GeV, and $N_{\phi}$ is 
a normalization factor
which we fix to be $0.76$ from the ATIC data at  low energies. 
The primary positron differential flux 
$\Phi^{\rm prim.}_{e^+}$ is $ (c/4\pi)f_{e^+}$,
where $f_{e^+}$ is given in (\ref{fe}).
Using  (\ref{fe})-(\ref{p3}), we then calculate appropriate 
quantities for PAMELA and ATIC:
\be
\frac{e^+}{e^+ +e^-} &=&
\frac{\Phi^{\rm prim.}_{e^+}+
\Phi^{\rm sec. bkg}_{e^+} }{\Phi^{\rm prim. }_{e^+}+
\Phi^{\rm sec. bkg}_{e^+}+\Phi^{\rm prim. bkg}_{e^-}
+\Phi^{\rm sec. bkg}_{e^-}}  ~~~\mbox{for PAMELA}, 
\label{pa}\\
E^3 \frac{dN }{d E}&=&E^3 (\Phi^{\rm prim. }_{e^+}+
\Phi^{\rm sec. bkg}_{e^+}+\Phi^{\rm prim. bkg}_{e^-}
+\Phi^{\rm sec. bkg}_{e^-})  
~\mbox{for ATIC}.
\label{at}
\ee
The results are shown in Figs.~\ref{pamela} and \ref{atic2},
where we have assumed $A_1=A_3=A_2$ in (\ref{a1}) and (\ref{a2}).
The blue lines are the predictions of the model, and 
we have used:~
$z=1/5\mbox{(dashed)}$, $1/6\mbox{(dot-dashed)}$, $1/5\mbox{(dotted)}$,
$\tau_{NDM}(0.11/\omega_{NDM} h^2)=1.44\mbox{(dashed)}$, 
$2.0\mbox{(dot-dashed)}$,
$3.0\mbox{(dotted)}~\times 10^{-26}~\mbox{sec}$,
$m_{NDM}=2.0\mbox{(dashed)}$, $1.5\mbox{(dot-dashed)}$, 
$1.0\mbox{(dotted)}~\mbox{TeV}$,
where $z$ is defined in (\ref{z}).
The predictions  are  written over the
figure 4 of the PAMELA paper \cite{pamela}
and the figure 3 of the ATIC paper \cite{atic}.
We see from Figs.~\ref{atic2} that
$m_{NDM}$ should be heavier than $O(1)$ TeV
in this model, too.

\section{Conclusion}
We have studied a dark matter model, in which one decaying and one stable
dark matter particles coexist. We have assumed that one of the 
discrete symmetries
ensuring the stability of the dark matter particles,
when imbedded into a larger group, is anomalous, and 
the heavier dark matter can decay non-perturbatively.
The huge suppression factor for the decay of dark matter
to be needed can be obtained in this way.
The concrete model we have considered 
is a supersymmetric extension of the Ma's inert Higgs model,
so that the decaying dark matter (the lightest right-handed neutrino)
can decay only into leptons along with  the stable dark matter (LSP).
We have shown that this scenario can explain 
the data of \cite{pamela, atic}.
It is clear that if the recent and future data coming 
from the cosmic ray observations are intimately related to 
the nature of dark matter, 
its explanation may open the window to new physics beyond the SM. 
The radiative dark matter decay  and high energy neutrino productions
will be our next projects.

\vspace*{5mm}
J.~K. is partially supported by a Grant-in-Aid for Scientific
Research (C) from Japan Society for Promotion of Science (No.18540257).
D.~S.. is partially supported by a Grant-in-Aid for Scientific
Research (C) from Japan Society for Promotion of Science (No.21540262).

\bibliographystyle{unsrt}

\begin{thebibliography}{99}



\bibitem{wmap}WMAP Collaboration, D.~N.~Spergel, {\it et al.},
	Astrophys. J. {\bf 148} (2003) 175; 
SDSS Collaboration, M.~Tegmark, {\it et al.}, 
Phys. Rev. {\bf D69} (2004) 103501.

\bibitem{oscil}SNO Collaboration, Q.~R~.Ahmad, {\it et al.},
	Phys. Rev. Lett. {\bf 89} (2002) 011301;
	Super-Kamiokande Collaboration, Y.~Fukuda, {\it et al.},
	Phys. Rev. Lett. {\bf 81} (1998) 1562; 
       KamLAND Collaboration, K.~Eguchi, {\it et al.}, 
       Phys. Rev. Lett. {\bf 90} (2003)
	021802; 
       K2K Collaboration, M.~H.~Ahn, {\it et al.},
	Phys. Rev. Lett. {\bf 90} (2003) 041801.    

\bibitem{susydm}For a review, see for example, 
G.~Jungman, M.~Kamionkowski and K.~Griest,
	Phys. Rept. {\bf 267} (1996) 195; G.~Bertone, D.~Hooper and
	J.~Silk, Phys. Rept. {\bf 405} (2005) 279.  


\bibitem{Ma:2006km}
  E.~Ma,
  Phys.\ Rev.\  D {\bf 73} (2006) 077301
  [arXiv:hep-ph/0601225].


\bibitem{seesaw} P.~Minkowski, Phys Lett. {\bf B67} (1977) 421;
	T.~Yanagida, in Proc. Workshop on Unified Theory and Baryon
	Number in the Universe, eds. O.~Sawada and A.~Sugamoto (KEK,
	1979); M.~Gell-Mann, P.~Ramond and R.~Slansky, in Supergravity,
	eds. P. van Nieuwenhuizen and D.~Freedman (North-Holland, 1979) p.315.

\bibitem{scdm}R.~Barbieri, L.~E.~Hall and V.~S.~Rychkov, Phys. Rev. {\bf
	D74} (2006) 015007; L.~Lepoz~Honorez, E.Nezri, J.~F.~Oliver and
	M.~H.~G.~Tytgat, JCAP {\bf 02} (2007) 28; M.~Gustafsson,
	E.~Lundstrom, L.~Bergstrom and J.~Edsjo, Phys. Rev. Lett. {\bf
	99} (2007) 041301.  

\bibitem{cdmmeg} J.~Kubo, E.~Ma and D.~Suematsu, Phys. Lett. 
{\bf B642} (2006) 18.



\bibitem{fcdm} L.~M.~Krauss, S.~Nasri and M.~Trodden, Phys. Rev. {\bf
	D67} (2003) 085002; D.~Aristizabal Sierra, J.~Kubo, D.~Restrepo,
	D.~Suematsu and O.~Zepata, Phys. Rev. {\bf D79} (2009) 013011;
	M.~Aoki, S.~Kanemura and O.~Seto, Phys. Rev. Lett. {\bf 102} 
(2009) 051805;  arXiv:0904.3829 [hep-ph].
    

\bibitem{ncdm} M.~Lattanzi and V.~W.~F.~Valle, Phys. Rev. Lett. {\bf 99}
	(2007) 121301; C.~Boehm, Y.~Farzan, T.~Hambye, S.~Palomares-Ruiz
	and S.~Pascoli, Phys. Rev. {\bf D 77} (2008) 043516; E.~Ma,
	Phys. Lett. {\bf B662} (2008) 49. 

\bibitem{ext} J.~Kubo and D.~Suematsu, Phys. Lett. {\bf B643} (2006) 336;
  Y.~Kajiyama, J.~Kubo and H.~Okada,
  Phys.\ Rev.\  D {\bf 75} (2007) 033001;
K.~S.~Babu and E.~Ma, Int. J. Mod. Phys. {\bf A23} (2008) 1813; 
D.~Suematsu, Eur. Phys. J. {\bf C56 } (2008) 379; E.~Ma and D.~Suematsu, 
Mod. Phys. Lett. {\bf A24} (2009) 583;
S.~Andreas, M.~H.~G.~Tytgat and Q.~Swillens,
  JCAP {\bf 0904} (2009) 004;
  D.~Suematsu, T.~Toma and T.~Yoshida,
  Phys. Rev. {\bf D79} (2009) 093004.



\bibitem{multidm}M.~Fairbairn and J.~Zupan, arXiv:0810.4147 [hep-ph].

\bibitem{Takayama:2000uz}
  F.~Takayama and M.~Yamaguchi,
  Phys.\ Lett.\  B {\bf 485} (2000) 388
  [arXiv:hep-ph/0005214].

\bibitem{Ibarra:2008jk}
  A.~Ibarra and D.~Tran,
  JCAP {\bf 0807} (2008) 002
  [arXiv:0804.4596 [astro-ph]];
  JCAP {\bf 0902} (2009) 021
  [arXiv:0811.1555 [hep-ph]].

   
\bibitem{Ishiwata:2008cu}
  K.~Ishiwata, S.~Matsumoto and T.~Moroi,
  Phys.\ Rev.\  D {\bf 78} (2008) 063505
  [arXiv:0805.1133 [hep-ph]];
  arXiv:0811.0250 [hep-ph];
  arXiv:0811.0250 [hep-ph];
  arXiv:0903.0242 [hep-ph].
  

  
  \bibitem{Yin:2008bs}
  P.~f.~Yin, Q.~Yuan, J.~Liu, J.~Zhang, X.~j.~Bi and S.~h.~Zhu,
  Phys.\ Rev.\  D {\bf 79} (2009) 023512
  [arXiv:0811.0176 [hep-ph]].
     
\bibitem{Hamaguchi:2008ta}
  K.~Hamaguchi, S.~Shirai and T.~T.~Yanagida,
  Phys.\ Lett.\  B {\bf 673} (2009) 247
  [arXiv:0812.2374 [hep-ph]].
  
  \bibitem{Nardi:2008ix}
  E.~Nardi, F.~Sannino and A.~Strumia,
  JCAP {\bf 0901} (2009) 043
  [arXiv:0811.4153 [hep-ph]].
  
  
\bibitem{Arvanitaki:2008hq}
  A.~Arvanitaki, S.~Dimopoulos, S.~Dubovsky, P.~W.~Graham, 
R.~Harnik and S.~Rajendran,
  arXiv:0812.2075 [hep-ph];
  arXiv:0904.2789 [hep-ph].
  
  \bibitem{Gogoladze:2009kv}
  I.~Gogoladze, R.~Khalid, Q.~Shafi and H.~Yuksel,
  arXiv:0901.0923 [hep-ph].
  
\bibitem{Shirai:2009fq}
  S.~Shirai, F.~Takahashi and T.~T.~Yanagida,
  arXiv:0905.0388 [hep-ph].
  

	
\bibitem{mindep}M.~Beltran, D.~Hooper, E.~W.~Kolb and Z.~A.~C.~Krusberg,
	arXiv:0808.3384 [hep-ph]; 
V.~Bager, W.-Y.Keung, D.~Marfatia and G.~Shaughnessy,
	Phys. Lett. {\bf B672} (2009) 141; I.~Cholis, L.~Goodenough,
	D.~Hooper, M.~Simet and N.~Weiner, arXiv:0809.1683 [hep-ph].

\bibitem{pamela} O.~Adriani {\it et al.}  [PAMELA Collaboration],
 Nature {\bf 458} (2009) 607
 [arXiv:0810.4995 [astro-ph]].

\bibitem{atic}J.~Chang {\it et al.}, Nature {\bf 456} (2008) 362.

\bibitem{Hisano:2003ec}
  J.~Hisano, S.~Matsumoto and M.~M.~Nojiri,
  Phys.\ Rev.\ Lett.\  {\bf 92} (2004) 031303
  [arXiv:hep-ph/0307216].

\bibitem{enhance}D.~Feldman, Z.~Liu and P.~Nath, arXiv:0810.5762; M.~Ibe,
	H.~Murayama, T.T.~Yanagida, arXiv:08120072.


\bibitem{positron}M.~Cirelli, M.~Kadastik, M.~Raidal and A.~Strumia,
	arXiv:0809.2409; Q.-H.~Cao, E.~Ma and G.~Shaughnessy,
	arXiv:0901.1334 [hep-ph].

\bibitem{Ma:2006uv}
  E.~Ma,
  Annales Fond.\ Broglie {\bf 31} (2006) 285
  [arXiv:hep-ph/0607142].	
  
    


\bibitem{Bi:2009md}
  X.~J.~Bi, P.~H.~Gu, T.~Li and X.~Zhang,
  JHEP {\bf 0904} (2009) 103
  [arXiv:0901.0176 [hep-ph]].
  
    \bibitem{Cao:2009yy}
  Q.~H.~Cao, E.~Ma and G.~Shaughnessy,
  Phys.\ Lett.\  B {\bf 673} (2009) 152
  [arXiv:0901.1334 [hep-ph]].
  
\bibitem{Chen:2009mf}
  C.~H.~Chen, C.~Q.~Geng and D.~V.~Zhuridov,
  arXiv:0901.2681 [hep-ph].





   
 
 \bibitem{Banks:1991xj}
T.~Banks and M.~Dine, Phys. Rev. \textbf{D45} (1992) 1424
  [hep-th/9109045].

\bibitem{Banks:1995ii}T.~Banks and M.~Dine,
Phys.\ Rev.\ D {\bf 50} (1994) 7454 [arXiv:hep-th/9406132];
Phys. Rev. \textbf{D53} (1996) 5790 [hep-th/9508071].
 
 \bibitem{ArkaniHamed:1998nu}
N.~Arkani-Hamed, M.~Dine, and S.~P. Martin, Phys. Lett. \textbf{B431} (1998)
  329 [hep-ph/9803432].

\bibitem{Ibanez:1991hv}
L.~E. Ib{\'a}{\~n}ez and G.~G. Ross, Phys. Lett. \textbf{B260} (1991)
  291.
 
 \bibitem{Araki:2006mw}
T.~Araki, Prog. Theor. Phys. \textbf{117} (2007) 1119
  [hep-ph/0612306].

\bibitem{Green:1984sg}
M.~B. Green and J.~H. Schwarz, Phys. Lett. \textbf{B149} (1984), 117--122.

\bibitem{Araki:2007ss}
  T.~Araki, K.~S.~Choi, T.~Kobayashi, J.~Kubo and H.~Ohki,
  Phys.\ Rev.\  D {\bf 76} (2007) 066006
  [arXiv:0705.3075 [hep-ph]].

\bibitem{Araki:2008ek}
  T.~Araki, T.~Kobayashi, J.~Kubo, S.~Ramos-Sanchez, M.~Ratz and 
P.~K.~S.~Vaudrevange,
  Nucl.\ Phys.\  B {\bf 805} (2008) 124
  [arXiv:0805.0207 [hep-th]].

\bibitem{Griest:1988ma}
  K.~Griest,
  Phys.\ Rev.\  D {\bf 38} (1988) 2357
  [Erratum-ibid.\  D {\bf 39} (1989) 3802].
  
\bibitem{Griest:1989zh}
  K.~Griest, M.~Kamionkowski and M.~S.~Turner,
  Phys.\ Rev.\  D {\bf 41} (1990) 3565; M.~Drees and M.~M.~Nojiri,
	Phys. Rev. D {\bf 47} (1993) 376. 
   
  \bibitem{Hisano:2005ec}
  J.~Hisano, S.~Matsumoto, O.~Saito and M.~Senami,
  Phys.\ Rev.\  D {\bf 73} (2006) 055004
  [arXiv:hep-ph/0511118].
  
   \bibitem{Moskalenko:1997gh}
  I.~V.~Moskalenko and A.~W.~Strong,
  Astrophys.\ J.\  {\bf 493} (1998) 694
  [arXiv:astro-ph/9710124].
  
    \bibitem{Delahaye:2007fr}
  T.~Delahaye, R.~Lineros, F.~Donato, N.~Fornengo and P.~Salati,
  Phys.\ Rev.\  D {\bf 77} (2008) 063527
  [arXiv:0712.2312 [astro-ph]].
  
  
  
  \bibitem{Navarro:1995iw}
  J.~F.~Navarro, C.~S.~Frenk and S.~D.~M.~White,
  Astrophys.\ J.\  {\bf 462} (1996) 563
  [arXiv:astro-ph/9508025].
 
  \bibitem{Baltz:1998xv}
  E.~A.~Baltz and J.~Edsjo,
  Phys.\ Rev.\  D {\bf 59} (1999) 023511
  [arXiv:astro-ph/9808243].
  


  
  \bibitem{Torii:2008xu}
  S.~Torii {\it et al.}  [PPB-BETS Collaboration],
  arXiv:0809.0760 [astro-ph].


  
\end{thebibliography}

\end{document}